\documentclass{article}

\usepackage{arxiv}
\usepackage[utf8]{inputenc} 
\usepackage[T1]{fontenc}    
\usepackage{hyperref}       
\usepackage{url}            
\usepackage{booktabs}       
\usepackage{amsfonts}       
\usepackage{nicefrac}       
\usepackage{microtype}      
\usepackage{lipsum}		
\usepackage{graphicx}
\usepackage{doi}
\usepackage{amsmath}
\graphicspath{ {./images/} }
\usepackage{subfigure}
\usepackage{placeins}
\usepackage{caption}

\title{Utilizing Segment Anything Model for Assessing Localization of GRAD-CAM in Medical Imaging}

\author{
  \textbf{Evan Kellener$^1$} \\
  University of California, Santa Cruz \\
  \texttt{ekellene@ucsc.edu}
  \and
    \textbf{Ihina Nath$^1$} \\
  University of California, Santa Cruz \\
  \texttt{inath@ucsc.edu}
  \and
    \textbf{An Ngo$^1$} \\
  University of California, Santa Cruz \\
  \texttt{andingo@ucsc.edu}
  \and
  \textbf{Thomas Nguyen$^1$} \\
  University of California, Santa Cruz \\
  \texttt{tnguy478@ucsc.edu}
  \and
  \textbf{Joshua Schuman$^1$} \\
  University of California, Santa Cruz \\
  \texttt{jschuman@ucsc.edu}
  \and
  \textbf{Coen Adler$^2$} \\ 
  University of California, Santa Cruz \\
  \texttt{ctadler@ucsc.edu}
  \and 
  \textbf{Arnav Kartikeya$^2$} \\
  University of California, Santa Cruz \\
  \texttt{akartike@ucsc.edu} 
}

\begin{document}

\def\thefootnote{1}\footnotetext{These authors contributed equally to this work}\def\thefootnote{\arabic{footnote}}
\def\thefootnote{2}\footnotetext{These individuals served as advisors on the paper}

\maketitle
\begin{abstract}
        The introduction of saliency map algorithms as an approach for assessing the interoperability of images has allowed for a deeper understanding of current black-box models with Artificial Intelligence. Their rise in popularity has led to these algorithms being applied in multiple fields, including medical imaging. With a classification task as important as those in medical domain, a need for rigorous testing of their capabilities arises. Current works examine capabilities through assessing the localization of saliency maps upon medical abnormalities within an image, through comparisons with human annotations. We propose utilizing Segment Anything Model (SAM) to both further the accuracy of such existing metrics, while also generalizing beyond the need for human annotations. Our results show both high degrees of similarity to existing metrics while also highlighting the capabilities of this methodology to beyond human-annotation. Furthermore, we explore the applications (and challenges) of SAM within the medical domain, including image pre-processing before segmenting, natural language proposals to SAM in the form of CLIP-SAM, and SAM accuracy across multiple medical imaging datasets.  
\end{abstract}

\section{Introduction}
Since the introduction of Convolutional Neural Networks (CNNs), its applications in image classification have spread throughout many domains. These include medical imaging, with tasks such as classifications of tumors, skin lesions, and diseases at the cellular level \cite{13, 14}. Given that CNNs are black-box models and considered not interpretable in their explanations for classifications \cite{15}, the need for an explanation of their decisions is a well-researched topic. Saliency maps provide such a post-hoc explanation for image classification tasks \cite{16}. GRAD-CAM is one of many popular methods for generating saliency maps on a trained model \cite{17}. It calculates a score for each pixel in an input image based on the gradients throughout the network that determine how important each pixel is for the output classification. This score for each pixel is later often interpreted through a heatmap superimposed upon the original input image.

Given the known inconsistencies with GRAD-CAM and other saliency-based XAI algorithms \cite{18}, there exists a need for assessing these saliency maps in their robustness. Previous works have attempted this topic, even within the medical domain \cite{9}. Current works use human-annotated bounded boxes to locate explanations for specific classifications within an image, then measure how well GRAD-CAM and other XAI algorithms localize upon those bounded boxes. We aim to generalize this work outside of the need for human-annotation, while also increasing the accuracy of a localization with segmentations instead of bounded boxes. In this report, we define a methodology for such an improvement, apply it to the case of identifying malaria within human blood smears, and compare results with current works. 

\subsection{Motivation for Metrics}
Due to the high-stakes nature of classification tasks within the medical field, it is essential that comparisons between saliency maps are quantified in order to draw accurate conclusions on the performance of such localization tools. Here we utilize the area under the precision-recall curve (AUPRC) \cite{9} metric as well as AUC-Judd metric to compare the various saliency methods using XAI techniques against human-annotated bounded-boxes. 

\section{Methodology}

\subsection{Data}
\begin{figure}[h]
    \centering
    \includegraphics[width=0.3\linewidth]{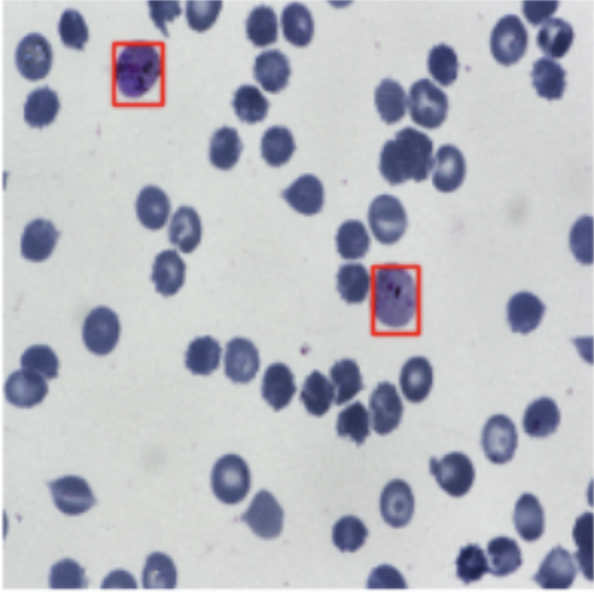}
    \caption{Manual bounded boxes around infected Malaria Cells}
    \label{fig:enter-label}
\end{figure}

For data to apply both SAM masks and GRAD-CAM saliency map explanations to, the Malaria Bounded Box dataset was chosen \cite{20}. Alternative datasets were considered although dismissed later due to issues with SAM, highlighted in the Challenges With SAM section further below. Furthermore, the human-annotation of the bounded boxes provide a ground-truth to compare our method against. The Malaria Bounded Box dataset consists of 1364 images of annotated human blood smears, spanning roughly 80,000 cells being analyzed. Cells are annotated and classified by type, including two classes of uninfected cells (RBCs and leukocytes) and four classes of infected cells (gametocytes, rings, trophozoites, and schizonts). For the purposes of a binary classification (the blood smear is infected or infected), all uninfected types were grouped as uninfected, and all infected were grouped as infected. Figure \ref{fig:enter-label} shows an annotated example of a singular blood smear, where annotated cells are infected.

\subsection{GRAD-CAM}
\begin{figure}[h]
    \centering
    \includegraphics[width=0.3\linewidth]{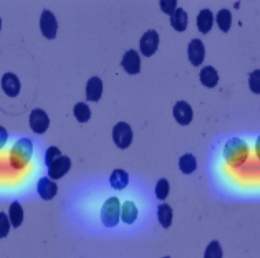}
    \caption{GRAD-CAM results on infected Malaria smear}
    \label{fig:gradcam-malaria}
\end{figure}

For producing saliency maps localizing upon malaria cells within an image, we use MobileNetv3 and a binary classification task in order to generate model for GRAD-CAM to utilize. MobileNetv3 was chosen for both performance reasons and GRAD-CAM results on classification being more accurate after rigourous testing with multiple ResNet models. For producing saliency map localizations around malaria cells, the task for MobileNet was a binary classification, determining whether a given blood smear contained infected cells or not. As discussed in the previous section, data was split into classes of infected or uninfected. MobileNetv3 is then fine-tuned on the split data with results of 96\% accuracy on a test split of data. Vanilla GRAD-CAM is then applied upon the fine-tuned model. Figure ~\ref{fig:gradcam-malaria} highlights GRAD-CAM superimposed upon a target image.

\subsection{Segment Anything Model}
In order to both increase the accuracy of annotation from bounded boxes while also removing the need for human-annotation, we attempt to use Segment Anything Model (SAM), a “foundational” model within image segmentation from Meta \cite{22}. SAM utilizes multiple encoder and decoder blocks for both feature extraction and prompt addition. These prompts exist in the form of points within the image, text, or bounded boxes. 

\subsection{Metrics}
To measure the accuracy of GRAD-CAM’s localizations based on either the masks generated by SAM or the human annotations of the malaria dataset, we use two metrics: AUC Judd and AUPRC.\\ \\
AUC, otherwise known as the Area Under the ROC Curve, is the most widely used metric for evaluating saliency maps \cite{7}. An ROC curve is plotted with the False Positive Rate as the x-axis and the True Positive Rate as the y-axis at different decision thresholds, and the area under the curve is the resulting value. Many variants of AUC exist for different applications, which differ in how the true and false positives are calculated - we chose to implement AUC Judd because it is a more consistent metric compared to other variants \cite{7}. In short, AUC Judd calculates true positives as predicted positive pixels at ground truth positive pixels, and false positives as the predicted positive pixels at ground truth negative pixels.\\ \\
Area Under the Precision-Recall curve (AUPRC) is less commonly used than AUC but is frequently used in medical applications because AUPRC is a more accurate metric for evaluating data with an imbalanced number of positive and negative labels \cite{8}. The Precision-Recall curve is plotted with recall on the x-axis, and precision on the y-axis, where recall is: \[\frac{\text{True Positives}}{\text{Total Actual Positives}}\] \\ and precision is: \\ \[\frac{\text{True Positives}}{\text{All Predicted Positives}}\]
\\ \\
Each metric has their strengths and weaknesses. AUC Judd is quickly informative of accuracy, but overlooks the effect of imbalanced positive and negative labels and as a result is overly optimistic at times \cite{8}. AUPRC is less intuitive to interpret because of a variable baseline value \cite{8} but represents the accuracy of a prediction more realistically on data with imbalanced positives and negatives, which is especially important for our data because there are much less pixels that make up malaria cells (positive labels) than non-malaria cells (negative labels). Nonetheless, both metrics are useful when evaluating data so we include both in our results.\\ \\
It is important to note when interpreting values that AUC and AUPRC have different baseline values, so it would not be fair to compare their values as is. The lowest possible value for AUC metrics are 0.5, while the lowest possible value for AUPRC is the ratio of positive labels to total number of labels \cite{8}. 

\section{Results}
After applying our methodology across 886 images (all images with infected cells), we find that SAM segmentations with a prompted bounded box are more accurate than just a bounding box, leading to better assessments of GRAD-CAM localization.

\begin{figure}[hb]
    \centering
    \includegraphics[width=0.5\linewidth]{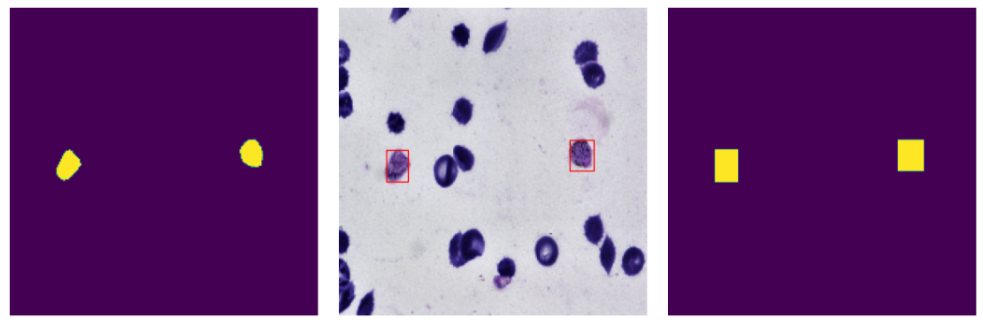}
    \includegraphics[width=0.5\linewidth]{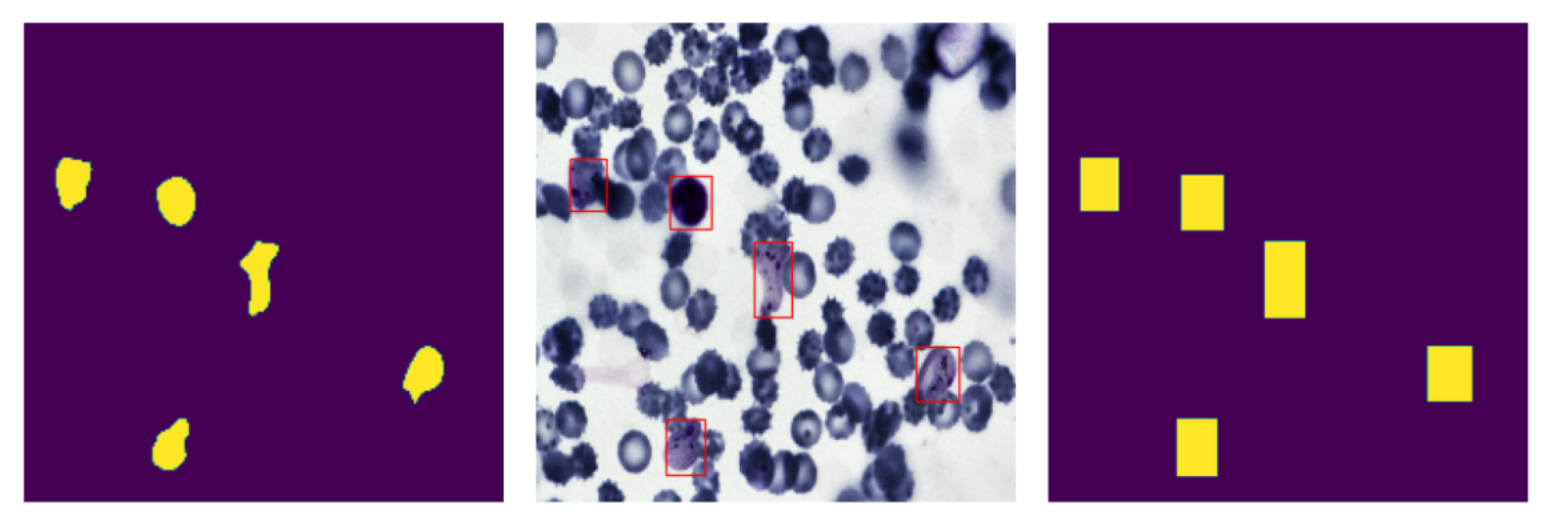}
    \caption{SAM segmentation given bounding box proposal}
    \label{fig:sam-vs-box}
\end{figure}


\begin{table}[!hb]
\centering
\begin{tabular}{ccc}
\hline
Metric & SAM & Bounded Box \\
\hline
Average AUPRC & 0.2212 & 0.2982 \\
\hline
Median AUPRC & 0.1752 & 0.2642 \\
\hline
Average AUC Judd & 0.7846 & 0.7832 \\
\hline
Median AUC Judd & 0.9206 & 0.9164 \\
\hline \\
\end{tabular}
\caption{AUPRC and AUC Judd statistics for SAM and Bounded Box}
\end{table}

\begin{figure}[!htb]
    \centering
    \begin{minipage}[b]{0.45\linewidth}
        \centering
        \includegraphics[width=\linewidth]{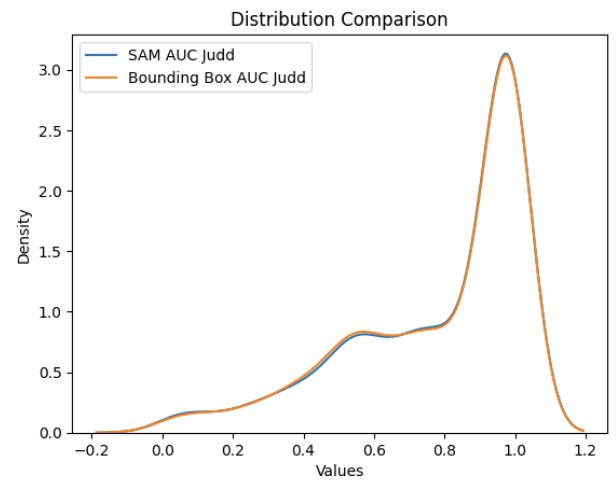}
        \label{fig:subfig1}
    \end{minipage}
    \hfill
    \begin{minipage}[b]{0.45\linewidth}
        \centering
        \includegraphics[width=\linewidth]{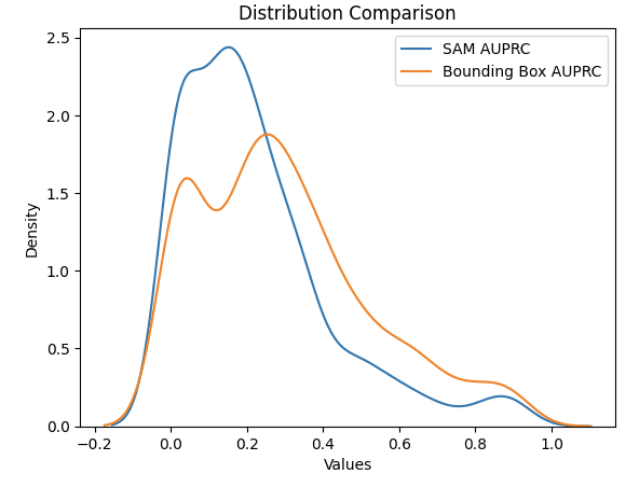}
        \label{fig:subfig2}
    \end{minipage}
    \caption{Comparison of AUC Judd/AUPRC score distribution between SAM and bounded box}
    \label{fig:graphs}
\end{figure}

AUC Judd between SAM and bounded box saliency maps when compared to GRAD-CAM are extremely similar. This suggests SAM segmentations have approximately the same/slightly better true positive to false positive ratio as annotated bounded boxes.  
Because of how lenient AUC Judd is, this indicates all SAM segmentations are contained within the bounding box, and in addition, they are not faulty.

The average AUPRC of SAM segmentations is lower than that of bounded boxes, which is expected when factoring in GRAD-CAM inconsistencies and the difference in area between a bounding box and a segmentation.
\begin{figure}[h]
    \centering
    \begin{subfigure}
        \centering
        \includegraphics{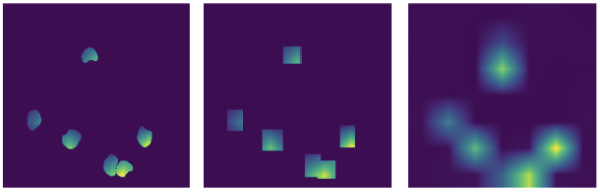}
        \caption*{AUPRC 0.23 vs 0.34}
        \label{fig:subfig1}
    \end{subfigure}%
    \vspace{0.5cm}
    \begin{subfigure}
        \centering
        \includegraphics{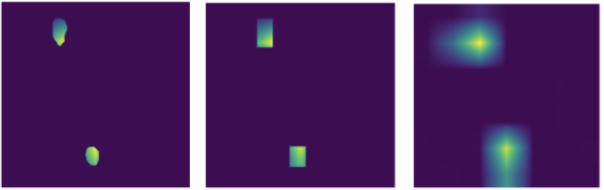}
        \caption*{AUPRC 0.50 vs 0.64}
        \label{fig:subfig2}
    \end{subfigure}
    \caption{Bounded box captures more GRAD-CAM inconsistencies (SAM has lower AUPRC)}
    \label{fig:saliencyimgs}
\end{figure}
We observe that SAM segmentations are much more rigorous in their assessment of GRAD-CAM compared to bounded boxes. SAM is less forgiving when GRAD-CAM is not within an area of interest, which shows in the contrast in average AUPRC (0.2212 vs 0.2982). In addition, bounding boxes inherently capture more of GRAD-CAM due to higher area, and assign a higher score even when GRAD-CAM results are clearly inaccurate. Figure ~\ref{fig:saliencyimgs} highlights such a discrepancy. 

\section{Challenges with SAM}
In our attempt to apply SAM to publicly available datasets, we encountered a series of challenges. Initial tests revealed SAM's difficulties in segmenting meaningful subsections of an image without proposals. One of these public datasets comprised brain scans with observable tumors, however, SAM was only capable of partially segmenting the tumors, and at times, failed to identify the tumors altogether. Analogous outcomes were observed in other public datasets involving cellular malaria and lung pneumonia \cite{12}.

\FloatBarrier
\begin{figure}[htbp]
  \centering
  
  \subfigure[Brain MRI scan that has a tumor]{\includegraphics[width=0.4\linewidth]{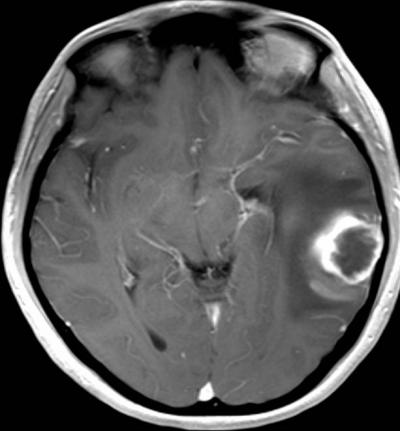}}
  \hfill
  \subfigure[Brain MRI scan segmented by SAM]{\includegraphics[width=0.4\linewidth]{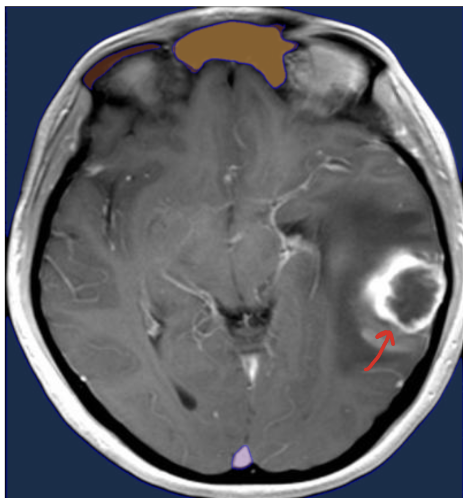}}
  
  \vspace{0.1cm} 
  
  \subfigure[Cell infected with malaria with no segmentation]{\includegraphics[width=0.4\linewidth]{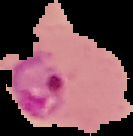}}
  \hfill
  \subfigure[Cell infected with Malaria segmented by SAM]{\includegraphics[width=0.4\linewidth]{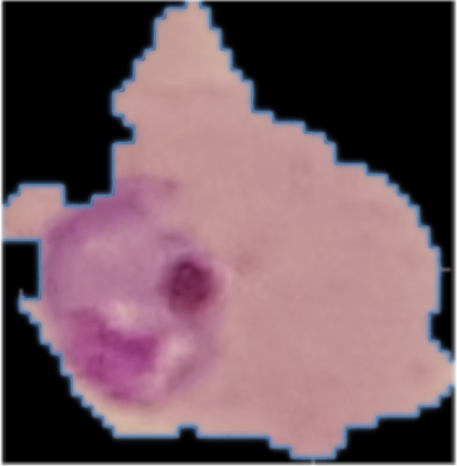}}
  
  \caption[Main Caption for the Whole Figure]{SAM failed to segment the brain tumor (a). Instead it segmented other parts of the brain (b), non related to the tumor. SAM segmenting the parasitized malaria data set tended to segment the entire cell (d) rather than the specifically highlighting the part of the cell containing malaria.}
  \label{fig:four-pictures}
\end{figure}

These findings echo the results from other studies. For instance, two separate investigations examining SAM's efficacy under 'everything mode' in segmenting camouflaged objects found its performance to be lacking \cite{1, 2}. In these situations, SAM struggled with visually complex scenes, such as animals camouflaged within their natural environments. Furthermore, SAM was found to be incapable of detecting concealed defects in industrial contexts \cite{2}. We suspect that due to similar reasons, SAM was unable to segment that medical imaging as well.

\subsection{Promptable SAM}
Notably, it has been suggested that using the promptable SAM yields more refined segmentation results, generating accurate masks tailored for specific uses \cite{3, 4}. With this in mind, we employed the prompt feature of SAM, providing illness-specific cues such as “malaria” in the malaria cell dataset or opting for more generic terms like “cell” in the same dataset. These experiments yielded varying results across the brain tumor, lung pneumonia, and malaria datasets, but none of the outcomes significantly improved the segmentation process. This disparity might be attributed to the possibility that SAM's training dataset lacked a substantial representation of medical imaging, thus providing no reference for identifying patterns indicative of diseases like malaria \cite{3}. In the future, testing with CLIP and SAM fine tuned upon medical datasets may yield better results.

\subsection{Image Preprocessing}
Seeking to enhance the accuracy of segmentations, we engaged in image preprocessing using two popular methods. One method we implemented involved initially defining a specific color range to construct a binary mask, with the intention of eliminating as many nonessential pixels as possible. Subsequently, we applied morphological operations to fill in gaps, smooth the boundaries of the region, and eliminate noise. To further enhance the image quality, we then increased the image contrast. Regrettably, this methodology resulted in substantial alterations to the dataset, which inadvertently led to a degradation in data quality. Furthermore, this approach necessitated additional effort in defining the specific color range for the pixels of interest. Ultimately, we decided that this method was to be abandoned as it was not practical in implementation.

\begin{figure}[htbp]
  \centering
  \includegraphics[width=0.6\linewidth]{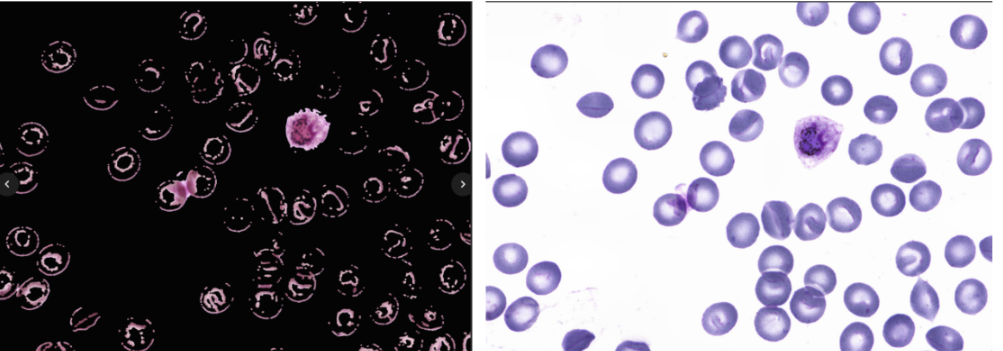}
  \caption{The preprocessed images depicted on the left exhibit a significant reduction in fine-grained details when juxtaposed with the original image showcased on the right. This particle image processing technique is considered to be suboptimal for medical imaging purposes, as the diagnostic process heavily relies on highly detailed visual information. The loss of such crucial details within the preprocessed images may lead to inaccurate diagnoses, showcasing the need to preserve the most detailed possible when preprocessing medical images. }
  \label{fig:my-picture}
\end{figure}

In our second method, we explored the idea of increasing the contrast of an image in order to further highlight the essential features for segmentation. This was tested on the Malaria Cell Images Dataset which consists of images of singular cells. The first method involved adaptive histogram equalization on the L-channel of the image which deals with the brightness. This method enhances brightness as well as contrast sensitivity. \\

\begin{figure}
    \centering
    \includegraphics[width=0.5\linewidth]{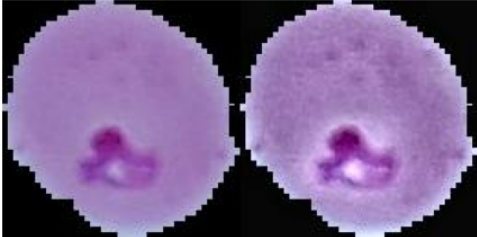}
    \caption{Original Cell Image (Left), Contrasted Cell Image (Right)}
    \label{fig:contrast-simple-cell}
\end{figure}

 \begin{figure}
     \centering
     \includegraphics[width=0.5\linewidth]{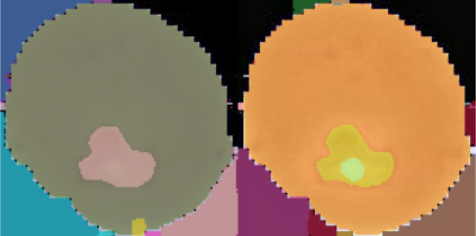}
     \caption{Original SAM Segmented Cell Image (Left), Contrasted SAM Segmented Cell Image (Right)}
     \label{fig:SAM-cell}
 \end{figure}

In Figure ~\ref{fig:my-picture}, the feature of interest i.e. malaria infected portion is clearly defined on the contrasted image as compared to the original.  \\ \\ Figure ~\ref{fig:SAM-cell} is the result of implementing SAM on the original cell image and on the modified contrasted image. Evidently, the image that has been modified to have higher contrast performed better as SAM was able to create a specific segmentation of the dark purple area that excluded the lighter portion in the middle. On the other hand, the entire area within the boundaries of the dark purple portion was segmented under one mask for the original image. This proved promising for more accurate segmentation of the malaria however it failed when tested against a more complicated cell image. \\ 
 \begin{figure}
     \centering
     \includegraphics[width=0.5\linewidth]{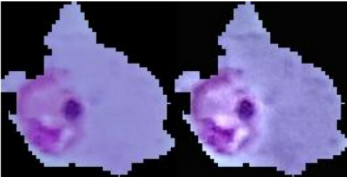}
     \caption{Original Complex Cell Image (Left), Contrasted Complex Cell Image (Right)}
     \label{fig:original-cell-contrast}
 \end{figure}

	The cell image in Figure ~\ref{fig:original-cell-contrast} is clearly more complex than the first example. Once again, the image on the right had the same enhancement to the brightness and contrast sensitivity. Unlike the first example, however, the modified image in this example did not perform as well as the modified image in the second example. \\

 \begin{figure}
     \centering
     \includegraphics[width=0.5\linewidth]{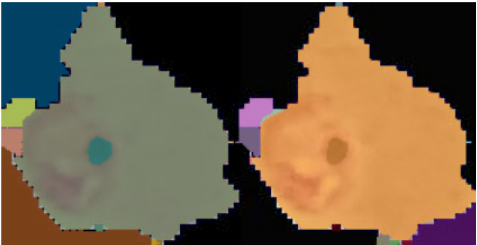}
     \caption{Original SAM Segmented Complex Cell Image (Left), Contrasted SAM Segmented Complex Cell Image (Right)}
     \label{fig:cell-SAM}
 \end{figure}

	Once SAM is implemented on both the original image and the modified image (Figure ~\ref{fig:cell-SAM}), it is clear that there is no significant improvement in segmentation.
In order to combat this problem, we applied modifications to this method. This once again involved adaptive histogram equalization on the L-channel but with additional amendments. Using OpenCV, weights were added to define the transparency/translucency of the new image. The brightness and contrast were further adjusted to isolate the point of interest. This was then applied to the more complex cell image as shown below:

\begin{figure}
    \centering
    \includegraphics[width=0.5\linewidth]{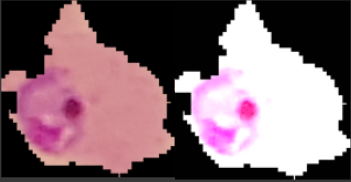}
    \caption{Original Complex Cell Image (Left), Complex Cell Image With Enhanced Contrast (Right)}
    \label{fig:contrast-cell}
\end{figure}

\begin{figure}
    \centering
    \includegraphics[width=0.5\linewidth]{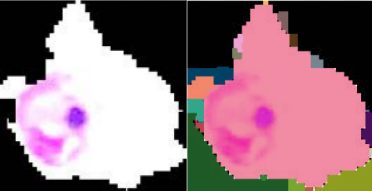}
    \caption{Complex Cell Image With Enhanced Contrast (Left), SAM segmented Complex Cell Image With Enhanced Contrast (Right)}
    \label{fig:contrast-cell-SAM}
\end{figure}

As seen in Figure ~\ref{fig:contrast-cell}, the original image has the purple area surrounded by pink while the modified image has only the affected area highlighted with the rest of the cell being white. Theoretically, SAM should have an easier time segmenting the modified image due to its stark contrast. However, SAM once again did not perform well in segmenting the affected area (Figure ~\ref{fig:contrast-cell-SAM}). It seemed to have segmented the entire cell and failed to differentiate the color.

\subsection{Integration with CLIP}
Despite these efforts, we encountered issues where SAM, even when successfully segmenting areas of interest, often generated multiple irrelevant or improper segmentations. A potential resolution we explored was integrating SAM with OpenAI's Contrastive Language–Image Pre-Training model (CLIP) \cite{5}. This model claims to have the ability to classify images accurately by using visual concepts from natural language supervision. Our new goal was to segment images using SAM, classify the resultant masks using CLIP, and then filter out irrelevant segmentations based on the classifications provided. Although CLIP demonstrated resilience in cases with limited reference data \cite{6} in other cases, the classifications it produced in medical imaging were inconsistent and inaccurate.

\subsection{Med-Clip Experiment}
Seeking an improvement over CLIP, we examined Med-Clip, a variant of CLIP explicitly trained on medical datasets. While Med-Clip appeared effective at classifying images within its training dataset, it struggled to accurately classify images outside of this set. We concluded that a more robust version of Med-Clip might enable the effective sorting of SAM-generated masks.

\FloatBarrier
\begin{figure}[htbp]
  \centering
  \begin{minipage}{0.3\linewidth}
    \centering
    \includegraphics[width=\linewidth]{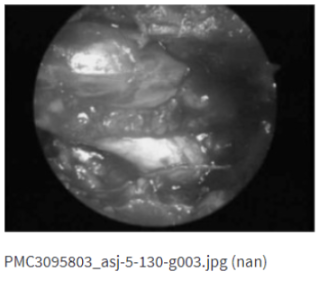}
  \end{minipage}%
  \hfill
  \begin{minipage}{0.3\linewidth}
    \centering
    \includegraphics[width=\linewidth]{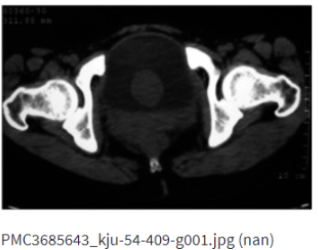}
  \end{minipage}%
  \hfill
  \begin{minipage}{0.3\linewidth}
    \centering
    \includegraphics[width=\linewidth]{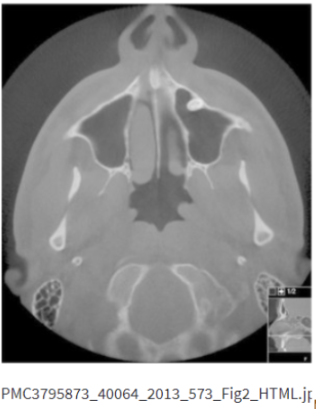}
  \end{minipage}
  \caption{When Med-Clip was prompted to provide images with malaria, it produced images unrelated to malaria or blood cells entirely.}
  \label{fig:three-pictures}
\end{figure} 

\subsection{Bounding Box Approach}
Ultimately, we leveraged SAM's bounding box feature, where a bounded box is provided to SAM that encapsulates all the pixels intended for segmentation. We used this approach with the malaria bounding box dataset since it already provided accurate bounding boxes, and SAM demonstrated a marked improvement in the accuracy of its segmentations. Not only did this resolve the issue of sorting the masks, but it also ensured the generated masks were only of malaria-infected cells. Furthermore, this gave us a ground truth to compare AUC Judd and AUPRC scores against.

\subsection{Conclusions Derived from the Challenges with SAM}
Our findings suggest that SAM's efficacy in medical image segmentation is largely contingent on the availability of pre-defined bounding boxes within the dataset. This highlights SAM's strong performance on discrete objects within a given region. Additionally, SAM's accuracy appears to be enhanced when the pathological features are highlighted in some manner.

\section{Future Works}

We propose an approach utilizing both the accuracy and generality of the SAM in applications of assessing saliency maps in the medical imaging domain. For the demonstration of our proposed methodology against a ground truth, and for reasons of challenges of SAM within the medical domain, our current work still utilizes human-annotated datasets. However, this approach with SAM will in theory be applicable to datasets without human annotation, and outside of the medical domain entirely. Future works can explore the application of the methods described in this paper within other domains and datasets without annotations. For further generalization to non-annotated datasets, customized pipelines or models may be necessary in conjunction with SAM. As discussed in the Challenges with SAM section, natural language as a proposal for SAM is neccesary for retrieving objects in question within an image requires the use of CLIP, which is not well defined for more specialized datasets such as those in the medical domain. Potential alternatives include utilizing a fine-tuned CLIP model as a prompt for SAM, potentially increasing the usefulness for medical datasets without human annotations. \\ \\ 
Furthermore, for the purposes of assessing saliency maps, comparisons against alternative XAI algorithms are necessary. Alternative approaches to saliency map explanations on imaging exist and can be tested, including Layer-wise Relevance Propagation and Guided GRAD-CAM\cite{22}, \cite{Selvaraju_2019}. 

\label{sec:others}









\bibliographystyle{unsrt}  
\bibliography{references}  
\end{document}